\documentclass[12pt]{article}
\usepackage{amsmath,amsfonts,amssymb,graphicx,epsfig}
\usepackage{colordvi}
\def\ba{\begin{eqnarray}}
\def\ea{\end{eqnarray}}

\begin{document}

\title{\bf Emission and absorption of photons and the black-body spectra in Lorentz-odd Electrodynamics}
\author{{{J.M. Fonseca}, A.H. Gomes, and {W.A.Moura-Melo}\thanks{E-mail: winder@ufv.br}} \\
\it \small \it Departamento de F\'{\i}sica, Universidade Federal
de Vi\c{c}osa\\ \small \it 36570-000, Vi\c{c}osa, Minas
Gerais, Brazil.}

\date{}

\maketitle
 \begin{abstract}
We study a number of issues related to the emission and absorption radiation by non-relativistic electrons within the framework of a Lorentz-breaking electrodynamics in (3+1) dimensions. Our main results concern how Planck-like spectrum law is sensitive to terms that violate Lorentz symmetry. We have realized that Planck law acquires extra terms proportional to the violating parameters: for the CPT-odd model, the leading extra terms appear to be linear or quadratic in these violating parameters according to the background vector is parallel or perpendicular to the photon wave-vector. In the CPT-even case a linear `correction' shows up. Among other possible ways to probe for these violations, by means of the present results, we may quote the direct observation of the extra contributions or an unbalancing in the mean occupation number of photon modes in a given thermal bath.
\end{abstract}

\section{Introduction and Motivation}
Symmetries are keystones for building the modern theories describing particle physics. As stated by the Noether theorem, continuous symmetries imply in dynamically conserved quantities, like energy-momentum and electric charge, whose conservation laws show up whenever the action remains invariant under space-time translations and local gauge transformations. Discrete symmetries are also very important in these frameworks, for instance, the combined CPT-invariance must be verified in all Lorentz-covariant local quantum field theories.\\

However, in the last years a number of effective models which do not accomplishes such a criterion has been proposed. There, theories are constructed to violate, for instance, Lorentz and CPT symmetries. Clearly, the acceptance of such proposals is deeply related to their theoretical consequences and/or observational demanding. Among such theories, those enclosed in the so-called Standard Model Extension\cite{Coll-KostPRD58-1998-116002} (SME) have received a great deal of efforts once its predictions upon symmetry violations are claimed to uncover new paradigms towards a {\em unified theory} which could consistently describes the quantum gravity. In the SME framework several non-standard coupling terms are allowed. Here, we shall consider two of them, responsible for Lorentz-breaking at the Abelian electrodynamic level, say in the radiation sector. One of these terms also violates CPT-operation, once it is parametrized by a constant vector-like background field which chooses a preferred direction in the space-time. Of course, such violations should be very small, once Lorentz and CPT symmetries have been confirmed to high precision in several experiments. Indeed, in the CPT-odd case, the background field magnitude is stringently constrained by astrophysical data, $|b_\mu|\lesssim 10^{-42}\,{\rm Gev}$ \cite{CFJ,KostMewesPRL99-011601-2007}; in turn, CPT-even may take larger values: suitable combinations of the dimensionless rank-4 tensor parameter could be around $10^{-16}$ \cite{KostMewesPRD66-2002-056005}.

Several works have been devoted to study how those extra terms modify conventional results concerning radiation and matter physical properties. For example, in the presence of the Lorentz- and CPT-odd and/or CPT-even terms a number of usual results concerning classical and quantum aspects of electromagnetic radiation acquire (small) contributions which often are linear or quadratic in the violating parameters. Among them are the Cerenkov \cite{Cerenkov,CerenkovCPTeven} and  synchrotron radiations\cite{syncroton-radiation}. Quantum mechanical effects could be also probed by means of two-level system \cite{ManoelPRD}. In turn, cosmic microwave background (CMB) data have been also investigated to search for possible traces of these violations in the very early Universe \cite{KostMewesPRL99-011601-2007,CMB-papers}. An even more amazing possibility is the photon splitting into two or more other on-shell photons within such frameworks\cite{photonsplitting}. Furthermore, there is also a extensive literature dedicated to study how such violations can be someway probed in the matter sector, dealing with electrons\cite{matter-electrons}, neutrinos\cite{matter-neutrinos}, and so forth. Many other aspects have been also extensively investigated, like dimensional reduction \cite{dimred}, causality and unitarity \cite{Adam-KlinkNPB607-2001-247,Adam-KlinkNPB657-2003-214}, and so on. However, additional results are important, for instance, to wide the possible experimental ways for probing such subtle symmetry breaking.\\

Here, we seek for possible effects of these violations in mechanisms of emission and absorption of quantum radiation by atoms. Namely, we realize that Planck law is sensitive to Lorentz breaking, accompanied by the CPT-odd or CPT-even terms. Although small, these deviations, linear or quadratic in the respective violating parameters, could be of prime importance once our results rely on mechanisms abundantly observed in nature. Indeed, the searching for those symmetry-breaking, based upon the present analysis (perhaps combined with others), include a very broad range of physical systems, from a relatively small number of atoms and photons to the CMB, which permeates the whole Universe. Namely, the observation of a tiny predominance of a given polarization over the another in a thermal bath could be taken as a good indication of such violations, as some of our results claim.\\

We organize our article as follows: in Section II, we present the model and some basic features of its associated radiation. Section III is devoted to the canonical quantization of the free radiation field. Indeed, only a brief account is given, sufficient for obtaining some useful results for later analysis. In the sequence, Section IV deals with the subject of emission and absorption of photons by non-relativistic atomic electrons, were among other results we find how Planck law reads within the frameworks with Lorentz-violation. Finally, we point out our Conclusions and Prospects for future investigation.\\

\section{The model and basic properties of its radiation}

Let us consider the Maxwell electrodynamics in (3+1) dimensions augmented by a CPT-odd and a CPT-even terms, both of them violating Lorentz symmetry, as follows\footnote{Our conventions read:
$\mu$,$\nu$, etc.=0,1,2,3, diag($\eta_{\mu\nu}$)=(+,-,-,-), and $\epsilon^{0123}=-\epsilon_{0123}=1$, etc. Natural units, with $\hbar=c=1$,
is used except where their presences are convenient.}:

\begin{equation}
{\cal L}_{\rm MED}=-\frac{1}{4} F_{\mu\nu}F^{\mu\nu}+\frac{1}{2}b_{\alpha}
A_{\beta}\tilde{F}^{\alpha\beta} -\frac14 d_{\mu\nu\alpha\beta} F^{\mu\nu}F^{\alpha\beta} \,,\label{Lmed}
\end{equation}
where $F_{\mu\nu}=\partial_\mu A_\nu -\partial_\nu A_\mu$ and $\tilde{F}^{\mu\nu}=\frac12 \epsilon^{\mu\nu\alpha\beta}F_{\alpha\beta}$. Sources can be introduced in the usual way, $A_\mu j^\mu$. Those terms proportional to the parameters\footnote{Instead of $b_\mu$ and $d_{\mu\nu\alpha\beta}$, it is more common to use $(k_{AF})_\mu$ and $(k_F)_{\mu\nu\alpha\beta}$, respectively. We justify our choosing for avoiding possible confusing of $k$ with the wave-vector label, $k_\mu$.} $b_\mu$ and $d_{\mu\nu\alpha\beta}$ are responsible for the violation of the Lorentz symmetry, but keeping gauge invariance under usual local transformations, $A_\mu(x) \,\to\, A_\mu(x) +\partial_\mu \Lambda(x)$. Their roles are distinct under CPT-operation: once $b_\mu$ is a constant background vector field, $\partial_\alpha b_\mu=0$, it implies in space-time anisotropy and ultimately in CPT-violation; contrary, $d_{\mu\nu\alpha\beta}$ respects this invariance. Additionally, it is dimensionless and bears the symmetries of the Riemann curvature tensor besides of having a vanishing double trace, so that only 19 components are independent. Another important difference between them lies on fact that while CPT-odd term yields a non-positive definite Hamiltonian, whenever $b_0\neq0$, all the CPT-even components leads to non-negative contributions to total energy (provided they are very small, as experiments strongly indicate; a good account on the latter is provided in Ref. \cite{KostMewesPRD66-2002-056005}). The dynamical equations for $A_\mu$ read:

\begin{equation}
\partial^\nu F_{\mu\nu} +d_{\mu\nu\alpha\beta}\partial^\nu F^{\alpha\beta} +b^\nu \tilde{F}_{\nu\mu}=0\,,\label{dyneqs}
\end{equation}
while the geometrical ones keep their usual form, $\partial_\mu \tilde{F}^{\mu\nu}\equiv 0$, stating the absence of magnetic sources. However, these exotic objects can be consistently introduced in this framework, provided that an extra electric current proportional to $b_\mu$ is induced\cite{nossoPRD2007}. From eq. (\ref{dyneqs}), the general dispersion relation may be obtained. However, treating each of the Lorentz-breaking term separately is more convenient and simpler. Thus, for the CPT-odd case (then, with $d_{\mu\nu\alpha\beta}=0$), we have:

\begin{equation}
(k_\mu k^\mu)^2 + (k_\mu k^\mu)(b_\nu b^\nu) -(k_\mu b^\mu)^2=0\,,\label{dispersionOdd}
\end{equation}
which is valid for arbitrary wave-vector, $k^\mu=(k^0, \vec{k})=(\omega,\vec{k})$, and $b_\mu$. The coupling between both vectors yields the splitting of the frequency modes and eventually to distinct phase velocities, even in vacuum (light birefringence phenomenon; further details below). On the other hand,  within the pure CPT-even framework (then, $b_\mu=0$) we find, to leading order:
\begin{equation}
\omega_{\pm}^{\rm even}= (1+\rho \pm \sigma) |\vec{k}|\,, \label{dispersionEven}
\end{equation}
where $2\rho=-\tilde{d}_\mu^{\;\;\mu}$ and $2\sigma^2= \tilde{d}_{\alpha\beta} \tilde{d}^{\alpha\beta} -2\rho^2$, with $\tilde{d}_{\mu\alpha}=d_{\mu\nu\alpha\beta} \,\hat{k}^\nu \,\hat{k}^\beta$ and $\hat{k}_\mu=k_\mu/|\vec{k}|$. Once these modes move at different phase velocities, $v_{\rm ph}=\omega/|\vec{k}|=c(1+\rho\pm \sigma)$, light experiences vacuum birefringence\cite{Coll-KostPRD58-1998-116002,KostMewesPRD66-2002-056005}.\\

Further features may also be worked out for the pure CPT-odd model. In this case, we have already mentioned that whenever $b_0\neq0$ a negative contribution to the Hamiltonian appears. Indeed, it was shown in Ref.\cite{Adam-KlinkNPB607-2001-247} that a consistent quantization of the radiation field is possible only for $b_\mu$ space-like, otherwise unitarity or causality is lost.\footnote{Indeed, unitarity or causality is lost in the pure time-like case, while it is kept if $b_\mu$ is pure space-like. In the light-like case there still lacks a complete analysis about the consistent quantization of the CPT-odd model, particularly, no definite answer has been given whether both unitarity and causality are preserved. Further details may be found, for instance, in Ref.\cite{Adam-KlinkNPB607-2001-247}.} Therefore, we take hereafter $b^\mu=(b^0;\vec{b})\equiv (0; m\hat{b})$, where $m$ is a parameter with mass dimension while $\hat{b}$ is a constant unity vector pointing along a preferred direction in the three-dimensional space. [In this case, the term in the Lagrangian is T-odd, C- and P-even, so that a direction in time is chosen by the model]. In this situation, the dispersion relation (\ref{dispersionOdd}) gets the form below:

\begin{equation}
(\omega_{\pm}^{\rm odd})^{2}=|\vec{k}|^{2}+\frac{1}{2}m^{2}\pm \frac{1}{2}m\sqrt{m^{2}+4(\vec{k}\cdot\hat{b})^{2}}\,. \label{dispersionOdd2}
\end{equation}
In general, these modes carry distinct mass-like gap proportional to $m$. For $\vec{k}\cdot\hat{b}=0$ one of the modes is massless while the another bears a mass-like gap equal to $m$. Note also that once their phase velocities are clearly different and depend on $m/|\vec{k}|$, they travel at distinct velocities even through vacuum. As we shall see, a number of results concerning emission and absorption of quantum radiation strongly depend on the frequency modes and how they couple to the wave-vector and other parameters, so that the dispersion relations (\ref{dispersionEven}) and (\ref{dispersionOdd2}) will be very important in our present work.\\

\section{Field quantization and basic results}

We now proceed to the canonical quantization of the radiation free field by expressing  $A_\mu$ in terms of plane-waves, inside a volume $V$. For that, we use the Coulomb gauge, $\nabla\cdot\vec{A}=0$ (no convenience seems to have in considering a covariant gauge, once Lorentz invariance is lost \cite{Adam-KlinkNPB607-2001-247}), so that:

\begin{eqnarray}
\label{Aexpansion}
\vec{A}(\vec{x},t)=\frac{1}{\sqrt{V}}\sum_{k}
         \left\{     \frac{1}{\sqrt{2\omega_{+}}}
\left[a_{+}(\vec{k})\,\vec{\varepsilon}_{+}(\vec{k})\,
e^{-i(\omega_{+}t-\vec{k}\cdot\vec{x})}\right]+
\frac{1}{\sqrt{2\omega_{-}}}\left[
a_{-}(\vec{k})\,\vec{\varepsilon}_{-}(\vec{k})\,
 e^{-i(\omega_{-}t\,+\vec{k}\cdot\vec{x})}\right]+ {\mbox{\rm H.C.}}    \right\}\,,
\end{eqnarray}
where H.C. accounts for the Hermitian conjugate terms. Note that expansion above is valid even in the general case of having both CPT-even and CPT-odd terms, like in Lagrangian (\ref{Lmed}), provided that the two remaining modes be taken into account or, alternatively, we write $\omega_\pm$ taking both contributions together. However, a clearer understanding of the physical contents and consequences of those terms for radiative processes is obtained by studying them separately. For example, the total energy and momentum carried by the radiation in the CPT-odd case, with $b^\mu=(0; m\hat{b})$, read as ${\bf H}_{\rm odd}= \sum_{\vec{k}}\, \left(\omega_+\, a^\dagger_+\, a_+  +  \omega_-\, a^\dagger_- \, a_-\right)$ and ${\vec{\bf P}}_{\rm odd} = \sum_{\vec{k}}\,\left(\vec{k}\, a^\dagger_+\, a_+  +  \vec{k}\, a^\dagger_-\, a_- \right)$. The counterparts for the pure CPT-even case may be worked out similarly. The creation and annihilation operators satisfy usual commutation relations, say:

\begin{equation}
\label{aadaggerccrs}
[a_{\pm}(\vec{k}),a_{\pm}(\vec{l}\,)]=[a^{\dag}_{\pm}(\vec{k}),a^{\dag}_{\pm}(\vec{l\,})]=0, [a_{\pm}(\vec{k}),a^{\dag}_{\pm}(\vec{l}\,)]=\delta_{+-}\delta_{\vec{k}\,\vec{l}}\,,
\end{equation}
where $\delta_{AB}$ is the Kronecker symbol (clearly, in the limit $V\to\infty$, we get $\delta_{\vec{k}\,\vec{l}}\,\to\, \delta^3(\vec{k}-\vec{l})$). Using these relations the Fock space may be build up as usual by imposing $a_+(\vec{k})| 0\rangle=a_-(\vec{k})| 0\rangle\equiv 0$. Therefore, whenever these operators act on general photon states we have:

\begin{equation}
\label{photonstates}
a^{\dag}_{\pm}(\vec{k})|n_{\vec{k},\pm}\rangle=\sqrt{n_{\vec{k},\pm}+1}|n_{\vec{k},\pm}+1\rangle \qquad {\mbox {\rm and}}\qquad a_{\pm}(\vec{k})|n_{\vec{k},\pm}\rangle=\sqrt{n_{\vec{k},\pm}}|n_{\vec{k},\pm}-1\rangle\,.
\end{equation}
For instance, $a^\dagger_+ (\vec{k})$ creates a photon with momentum $\vec{k}$ and frequency mode $\omega_+$, while $a_+ (\vec{k})$ annihilates it. Let us recall that the relations between frequency and momentum are given by eqs. (\ref{dispersionEven}) and (\ref{dispersionOdd2}), according to the pure CPT-even or CPT-odd (with $b^0=0$ and $\vec{b}=m\hat{b}$) frameworks. More general states, including distinct polarizations, may be worked out by direct product, say, $|n_{\vec{k}_{1},+};\, m_{\vec{k}_{2},-};\,\ldots \rangle = |n_{\vec{k}_{1},+}\rangle \otimes | m_{\vec{k}_{2},-}\rangle\otimes\,\ldots$. The photon number operator, $N_{\vec{k}, \pm}=a^\dagger_+ a_+ +a^\dagger_- a_-$,
its basics properties and uncertainty relations may be worked out. For example, the uncertainty product of the number and phase operators reads $\Delta N\, \Delta \phi>1$, as usual. \\

For the sake of completeness, we also discuss about the explicit expressions for the polarization operators which strongly depend upon the relation between $\omega$ and $\vec{k}$. For the CPT-odd case, they get the forms below:

\begin{eqnarray}
\label{polarizationsCPTOdd}
\vec{\varepsilon}_{\pm}(\vec{k})=\frac{1}{\sqrt{2\Gamma(\Gamma \pm m)}}[2\omega_{\pm}\cos\theta\hat{\xi}_{1}\mp i(\Gamma \pm m) \hat{\xi}_2]\,,
\end{eqnarray}
so that $(\hat{\xi}_{1},\hat{\xi}_{2},\hat{k})$ form a right-handed orthonormal basis, and $\Gamma=\Gamma(\vec{k})\equiv\sqrt{m^{2}+4(\vec{k}\cdot\hat{b})^{2}}$. For the CPT-even case we also obtain\cite{KostMewesPRD66-2002-056005}:
\begin{equation}
\vec{\varepsilon}_{\pm}(\vec{k})=\frac{1}{\sqrt{1\mp 2 \cos\theta}} \left[ \sin\theta\, \hat{\xi}_1 \pm (1-\cos\theta)\hat{\xi}_2 \right]\,,
\end{equation}
with $\tan\theta=2\tilde{d}^{12}/ (\tilde{d}^{11} - \tilde{d}^{22})$. Namely note that   $\vec{\varepsilon}_{\pm}\cdot \hat{k}=0$ and $\vec{\varepsilon}_{\pm}\cdot\vec{\varepsilon}_{\pm}=\delta_{+-}$.

\section{Emission and absorption of photons in a Lorentz-odd framework}

For analyzing some features of the electromagnetic radiation emerging from these Lorentz-violating frameworks, we consider the case of emission and absorption of photons by non-relativistic atomic electrons. Explicitly, we shall consider the case where an atom begins at a quantum state $A$, interacts with a photon, characterized by $(\omega_+, \vec{k})$, and ends at $B$ (the transition involving a mode with $\omega_-$ gives analogous results). At first order, the absorption process is described by:

\begin{equation}
\label{ABtransition}
\langle B;n_{\vec{k},+}-1|H_{I}|A;n_{\vec{k},+}\rangle\,.
\end{equation}
The interacting Hamiltonian, $H_{I}$, acts on atomic states $A$ and $B$, as well as on photon states (indeed, a complete state is the direct product of the atom and photon states, as above), and is written as $H_I=-\frac{e}{m_e} \sum_i\, \vec{A}(\vec{x}_i,t)\cdot \vec{p}_i$, where $e$ and $m_e$ are the electronic charge and mass, while $\vec{A}(\vec{x}_i,t)$ is the vector potential, at time $t$, acting on an electron with (non-relativistic) momentum $\vec{p}_i$ placed at $\vec{x}_i$. Taking eq. (\ref{Aexpansion}) to the transition above we get, after some algebra:

\begin{eqnarray}
\label{ABabsorption}
\langle B;n_{\vec{k},+}-1|H_{I}|A;n_{\vec{k},+}\rangle=\nonumber \\ -\frac{e}{m}\sqrt{\frac{n_{\vec{k},+}}{2V\omega_{+}}}\sum_{i}\langle B|e^{i\vec{k}\cdot \vec{x}_{i}}\vec{p}_{i}\cdot \vec{\varepsilon}_{+}|A\rangle\, e^{-i\omega_{+}t}\,.
\end{eqnarray}
As expected from common experience, it is an impossible event the absorption of a photon with polarization distinct from that released from the electromagnetic field (such a result is related to the orthogonality of photon states, expressed by the commutation relations between $a$ and $a^\dagger$, eq. (\ref{aadaggerccrs})). In general, a process in which an initial state goes to a final one with different photon polarization is forbidden. Analogously, if the atom emits  a photon, the transition reads:

\begin{equation}
\label{ABemission}
\langle B;n_{\vec{k},+}+1|H_{I}|A;n_{\vec{k},+}\rangle= -\frac{e}{m}\sqrt{\frac{(n_{\vec{k},+}+1)}{2V\omega_{+}}}\sum_{i}\langle B|e^{-i\vec{k}\cdot \vec{x}_{i}}\vec{p}_{i}\cdot \vec{\varepsilon}^{\ast}_{+}|A\rangle e^{i\omega_{+}t}\,.
\end{equation}
In form, the expressions above are identical to the usual ones. Nevertheless, the dispersion relations are now different from the standard case, $\omega=|\vec{k}|$, so that several novelties will show up, as below.\\

Now, suppose a number of such atoms interacting with a radiation field by means of reversible processes, $A \rightleftharpoons B+\gamma$, which maintain thermal equilibrium. Thus, if the population number of initial state is $N(A)$ and of the final state is $N(B)$, then we have, at thermal equilibrium:

\begin{equation}
\frac{N(B)}{N(A)}= \frac{e^{-E_B/k_B T}}{e^{-E_A/ k_B T}}=e^{\hbar \omega/k_BT}\,,
\label{NANB}
\end{equation}
where $E_A, E_B$ represent the total photon energy of states $A$ and $B$, respectively, while $k_B$ is the Boltzmann constant. The expression above may be rewritten as $N(B){\cal P}_{\rm abs}=N(A) {\cal P}_{\rm emis}$, where ${\cal P}_{\rm abs}, {\cal P}_{\rm emis}$ are the transition probabilities associated to the processes $B+\gamma\longrightarrow A$ (absorption) and $A\longrightarrow \gamma +B$ (emission), respectively. At leading order, general time-dependent perturbation gives:

\begin{equation}
{\cal P}=\frac{2\pi}{\hbar} |\langle m|H_I|l \rangle|^2 \, t\, \delta(E_m -E_l \pm \hbar \omega) \,,\label{transproba}
\end{equation}
with the transition $l\to m$ taking place by emitting (+) or absorbing (-) a photon of frequency $\omega$. Now using eqs. ({\ref{ABabsorption}) and (\ref{ABemission}), we get (analogous expressions read for polarization $\omega_-$ and for the CPT-even case, as well):

\begin{equation}
\frac{{\cal P}_{\rm emis}}{{\cal P}_{\rm abs}}= \frac{(n_{\vec{k},+}+1)\sum_{i}\, |\langle B|e^{-i\vec{k}\cdot \vec{x}_{i}} \, \vec{p}_{i}\cdot \vec{\varepsilon}^{\;*}_{+}|A\rangle|^{2}}{n_{\vec{k},+}\sum_{i}\,|\langle A|e^{i\vec{k}\cdot \vec{x}_{i}}\, \vec{p}_{i}\cdot \vec{\varepsilon}_{+}|B\rangle|^{2}}\, , \label{PemPab}
\end{equation}
which immediately yields $n_{\vec{k},+} N(B)=(n_{\vec{k},+} +1) N(A)$. Now, taking this result to eq. (\ref{NANB}), we readily obtain (for $\omega_-$, analogous result follows in the same way):

\begin{equation}
n_{\vec{k},\pm}= \frac{1}{e^{\hbar \omega_\pm/k_B T}-1}\,, \label{nkthermal}
\end{equation}
which gives the average number of photons at each state characterized by $(\omega_+;\vec{k})$ or $(\omega_-;\vec{k})$. Thus the average energy by photon state reads $\overline{E}_\pm=\hbar \omega_\pm n_{\vec{k}, \pm}$. These results, together with the dispersion relations (\ref{dispersionEven}) and (\ref{dispersionOdd2}), clearly show that the radiation field is homogeneous, once the quantities above do not depend on vector position, for both CPT-even and CPT-odd frameworks. However, such a field is anisotropic in the CPT-odd case, since it depends on the relative orientation of $\vec{k}$ and the background field, $\vec{b}$, as expressed by eq. (\ref{dispersionOdd2}). In addition, in both cases each polarization state experiences Lorentz anisotropy differently, say, at a given temperature, $T$, and momentum, $\vec{k}$, we generally have $n_{\vec{k}, +}\neq n_{\vec{k},-}$.\\

As in the usual case, eq. (\ref{nkthermal}) does not depend on the geometrical details of the system neither on its internal properties, so that even in these cases where Lorentz symmetry is lost, we can still speak of universal spectrum of emission and absortion of radiation. Therefore, two different systems may achieve thermal equilibrium through sucessive interchanging of photons. Thus, the usual equivalence between cavity and black-body radiations seems to be kept.\\

Now, assuming usual periodic boundary conditions, the momenta of the radiation field enclosed in a cubic volume $V=L^3$ are given by $k_i= 2\pi n_i/L$ where $i=1,2,3=x,y,z$ and $n_x,n_y,n_z=\pm1,\pm2,\pm3\ldots$. Then, the total number of quantum radiation oscillators (photons) with polarization $\omega_+$, energy between $[\hbar \omega_+, \hbar(\omega_+ +d\omega_+)]$, and propagating towards a direction enclosed by a solid angle $d\Omega$ equals the volume element in the three-dimensional $n$-space, $n^2 dn \, d\Omega=\rho_{\omega,d\Omega}dE$ \cite{Heitler,Sakurai}. In the latter equality $\rho_{\omega,d\Omega}={\cal N} \Pi_i dk_i$ is the so-called density of allowed states per unity frequency, $\omega$, while ${\cal N}$ is the number of polarizations. Therefore, energy density per unity frequency (and per polarization) is given by the total energy enclosed in the volume times the density of states, $\int_V \rho_{\omega,d\Omega}dE$. Evaluating this expression we obtain (explicitly $c$ and $\hbar$):

\begin{equation}
u(\omega_\pm)= \frac{4\pi}{(2\pi)^{3}}\frac{\hbar\omega_\pm}{e^{\hbar\omega_\pm/k_B T}-1}\, \frac{k^{2}dk}{d\omega_\pm} \,. \label{U+-}
\end{equation}
The total energy density per frequency, $u(\omega)$, in a given radiation thermal bath is given by summing over the respective contributions from each polarization. Although expression above has the usual form, it should be emphasized that differences actually appear once that the evaluation of the term $\omega_{\pm}k^{2}dk/d\omega_\pm$ is strongly dependent on the dispersion relations, as follows.\\

Firstly, let us consider the CPT-odd framework. Using dispersion relation (\ref{dispersionOdd2}) it is easy to show that (at regimes where $\omega>>\omega_0$):

\begin{equation}
\frac{\omega_\pm k^2 dk}{d\omega_\pm} = \frac{\omega^{3}}{c^{3}}\Big(1\pm\frac{\omega_{0}\cos\theta}{2\omega}+ \frac{\omega_{0}^{2}}{4\omega^{2}}(2-\cos^{2}\theta)\Big)+{\cal O}(\omega^3_0)\,,
\end{equation}
where $\omega=\omega(k)=c|\vec{k}|$ and $\omega_0\equiv mc^2/\hbar$ are the dynamical and rest-like ($k$-independent) frequencies. The non-vanishing $\omega_0$ is related to the mass-like gap previously discussed which implies in a sort of rest energy for electromagnetic radiation in this case.The angle $\theta$ lies between the vectors $\vec{k}$ and $\vec{b}$. Now, computing the energy density per frequency we get (up to order $\omega^2_0$):

\begin{equation}
u^{\rm odd} (\omega_{\pm},T)|_{\omega>>\omega_0}= \frac{4\pi\hbar}{(2\pi c)^3}\frac{\omega^3}{e^{\hbar\omega/k_B T}-1} \left(1\pm\frac{\omega_{0}\cos\theta}{2\omega}+ \frac{\omega_{0}^{2}}{4\omega^{2}}(2-\cos^{2}\theta)\right),\, \label{Uoddgeneral}
\end{equation}
from what there follows the energy density per polarization mode (thereof, the pre-factor $1/2$ below), $U_\pm=\int^\infty_0 u(\omega_\pm,T) d\omega_\pm$:

\begin{equation}
U_\pm(T)=\frac12\sigma_{0}T^{4}\Big[1\pm \sigma_{1}\, \omega_0\cos\theta\, T^{-1} +
\sigma_{2}\, \omega^2_0(2-\cos^{2}\theta)T^{-2}\Big]\,,
\end{equation}
where $\sigma_{0}=\pi^{2}k_{B}^{4}/15(\hbar c)^{3}\approx 7.56\times10^{-16} {\rm J/m^{3}K^{4}}$ is the Stefan-Boltzmann constant, while $\sigma_{1}\equiv  15\hbar \zeta(3)/\pi^4 k_B\approx1.41\times10^{-12} {\rm K.s}$ ($\zeta(3)\approx 1.2$, $\zeta(x)$ is the $\zeta$-Riemann function) and $\sigma_{2}\equiv \,5\hbar^2/\pi^2\, k^2_B\approx 3.69\times10^{-41}\, {\rm K^{2}.s^2}$.  Recalling that\cite{KostMewesPRL99-011601-2007} $m\lesssim 10^{-42} {\rm Gev}/c^2$, what gives $\omega_0\lesssim 10^{-17}\, {\rm Hz}$, then $\sigma_1 \omega_0\approx 10^{-30} \,{\rm K}$ which is negligible comparared to the unity, in equation above. However, considerable compensation may come from a very low temperature, say, $T\sim 10^{-10} \,{\rm K}$ (already achieved in laboratories), so that the leading `correction' to Stefan-Boltzmann law associated to Lorentz and CPT violations is around $10^{-20}$. Black-body-type radiation has been also studied in non-commutative geometry frameworks. There, additional `contributions' appear proportional to higher powers of $T$, $T^8$ at leading order, so contrary to the our case, most probably probed at very high temperature \cite{black-body-NCG}. Similar results to those above have been also obtained in the work of Ref.\cite{Casana}\\

For the sake of clarity it may be interesting to compute how Planck law is modified for two special cases of $\vec{k}\cdot\vec{b}$. Let us start from that where the photon momentum is parallel to the violating vector, $\vec{k}\cdot \hat{b}=k$. Without loss of generality let us take $\vec{k}=(0,0, k=|\vec{k}|)$ and $\hat{b}=(0,0,1)$, yielding $4\omega^2_\pm=c^2[\sqrt{4k^2 +m^2} \pm m]^2 $, so that, after some algebra we obtain:

\begin{equation}
u^{\rm odd}_{\parallel} (\omega_{\pm},T)|_{\omega>>\omega_0}= \frac{4\pi\hbar}{(2\pi c)^3}\frac{\omega^3}{e^{\hbar\omega/k_B T}-1} \left(1 \pm \frac12 \frac{\omega_0}{\omega} +\frac14 \frac{\omega^2_0}{\omega^2} +{\cal O}(\omega^3_0)\right)\,. \label{Uoddpar}
\end{equation}
Note that the leading deviation appears linearly in the violating parameter and increases(decreases) the density energy per frequency according to polarization is positive (negative, $\omega_-$). Although very small, the correction above could imply in a (small) unbalancing between the polarizations in thermal equilibrium, as already indicated by eq.(\ref{nkthermal}). Indeed, in evaluating the total energy density per (dynamical) frequency, $E(\omega)$, we have two possibilities, each of them leading to a different way to probe for these violations. First, if we assume that energy is equally partitioned into the two modes at each frequency, say $E(\omega)=E(\omega_+) + E(\omega_-)$, then the linear contributions from $E^{\rm odd}_{\parallel} (\omega_{\pm})$ above identically cancel each other, and the correction appears only at $\omega^2_0$, and positive. On the other hand, the number of photons associated to $\omega_-$ should be higher by a factor linear in $\omega_0$, once frequency and energy are linearly related. This is the case we would expect to be most physically plausible, at least when Lorentz and CPT, among other symmetries, are respected. Second, suppose that the number of photons with both polarizations equal at a given dynamical frequency, $\omega=\vec{k}$. In this case, the total energy per frequency would be $E(\omega)=\alpha E(\omega_+) + \beta E(\omega_-)$, so that a net linear correction in $\omega_0$ would be obtained for $E(\omega)$, once $\alpha\lesssim\beta$ so that $\alpha+\beta=1$. The analysis and discussion above are strictly valid for usual radiation, say, with $\omega\gtrsim$ some powers of Hz (say, $\omega>>\omega_0$), but if we could probe radiation with very small frequencies, ideally compared to $\omega_0$, then things should change and a trace of Lorentz and CPT violations could be clearer probed. Actually, for $\omega_0\approx \omega$, we obtain the analogue of the Rayleigh-Jeans classical result:
\begin{equation}
u^{\rm odd}_{\parallel} (\omega_{\pm},T)_{\omega\approx\omega_0}\simeq \frac{4\pi}{(2\pi c)^3}\frac{2 k_B T\omega^2}{(\sqrt5 \pm 1)}\,,
\end{equation}
which clearly indicate that at such regimes energy difference between the modes cannot be neglected. Moreover, we should stress that, as $\vec{k}\to0$ ($\omega\to0$) then $u(\omega)$ identically vanishes, stating that, even in this case where the modes bear rest-like frequencies, non-vanishing thermal equilibrium is achieved by means of radiation dynamics, whenever interacting with (non-relativistic) electrons.\\

Let us now consider the case where the photon momentum is perpendicular to the background vector, say, $\vec{k}\cdot \hat{b}=0$, so that the dispersion relation is now simplified to $2\omega^2_\pm=[2k^2 + m^2(1\pm1)]$. The remaining details may be worked out as above yielding, for arbitrary $\omega$:

\begin{equation}
u^{\rm odd}_{\perp}(\omega_\pm,T)= \frac{4\pi\hbar}{(2\pi c)^3}\frac{\omega^3}{e^{\hbar\omega_\pm/k_B T}-1} \left(1 +\frac12 \frac{\omega^2_0}{\omega^2}\right)\,, \label{Uoddperp}
\end{equation}
whose correction comes only at second order, $\omega^2_0$, for any finite value of $\omega$, so falling off as $\vec{k}\to0$. Thus, if we probe for Lorentz and CPT violations by means of a thermal bath of photons it would be more suitable to do that where radiation travels along the background vector direction, for what the deviations appear linearly in $\omega_0$.\\

Now, let us carry out the CPT-even case, so that the modes are given, at leading order, by $\omega_\pm=c|\vec{k}|(1+\rho\pm\sigma)= \omega(1+\rho\pm\sigma)$. After some algebra the Planck-like law is obtained to be:

\begin{equation}
u^{\rm even}(\omega_\pm,T)= \frac{4\pi\hbar}{(2\pi c)^3}\frac{\omega^3}{e^{\hbar\omega_\pm/k_B T}-1}\approx \frac{4\pi\hbar}{(2\pi c)^3}\frac{\omega^3}{e^{\hbar\omega/k_B T}-1} \left(1-\frac{\hbar \omega}{k_B T} (\rho \pm \sigma) +{\cal O}((\rho\pm \sigma)^2)\right) \,.\label{Ueven}
\end{equation}
What clearly shows that the extra contribution associated to the CPT-even parameter comes linearly. This may become even more interesting if we sum up over the modes with the assumption of equally partitioned energy, $E(\omega)=E(\omega_+)+ E(\omega_-)$, so that:
\begin{equation}
 u^{\rm even}(\omega,T)= \frac{8\pi\hbar}{(2\pi c)^3}\frac{\omega^3}{e^{\hbar\omega/k_B T}-1} \left(1-\frac{\hbar \omega}{k_B T}\rho \right)\,.
\end{equation}
Then, besides a difference in the number occupation of the modes, linear in $\rho$, we have also obtained a density energy per frequency which bears a linear extra term, $-\frac{\hbar \omega}{k_B T}\rho$ (note that in a similar situation, the CPT-odd case, with $\omega>>\omega_0$, the deviation is proportional to $+\omega^2_0$). For instance, for CMB radiation ($T\sim 3{\rm K}$, $\omega\sim 10^{11}\,{\rm Hz}$) we get $\frac{\hbar \omega}{k_B T}\rho\sim \rho$. Therefore, if $\rho$ is constrained to be around $10^{-16}$, the latter expression predicts a small deviation in the CMB power spectra around this value. Other effects could be somewhat combined altogether in order to enhance the chances to detect any trace of Lorentz symmetry breaking or to set new and more precise bounds on this parameter.\\

Before closing this section, let us recall that a semi-classical analysis is always possible for both emission and absorption processes regarding the number of photons is sufficiently large (radiation is intense enough). Indeed, according this description the potential is given by:

\begin{equation}
\vec{A}^{(abs)}=\vec{A}^{(abs)}_{0}e^{i\vec{k}\vec{x}-i\omega_{+}t},\;\;\;\;\;
\vec{A}^{(abs)}_{0}=\sqrt{\frac{n_{\vec{k},+}}{2V\omega_{+}}}\vec{\varepsilon}_{+},
\end{equation}
\begin{equation}
\vec{A}^{(emis)}=\vec{A}^{(emis)}_{0}e^{-i\vec{k}\vec{x}+i\omega_{+}t},\;\;\;\;\;
\vec{A}^{(emis)}_{0}=\sqrt{\frac{(n_{\vec{k},+}+1)}{2V\omega_{+}}}\vec{\varepsilon}_{+}~,
\end{equation}
from what the transition rates may be obtained and show to equal those previously evaluated, eqs. (\ref{ABabsorption}) and (\ref{ABemission}). The main advantage of this approach is that it is valid for any choice of $b_\mu$, including $b_0\neq0$ or even pure time-like. Now, if we assume that detailed balance holds (at thermal equilibrium each photon emission must be counterbalanced by its equivalent absorption process), $N(B){\cal P}_{\rm abs}=N(A) {\cal P}_{\rm emis}$, then those fundamental results there follow, like eqs. (\ref{NANB}) and (\ref{nkthermal}).

To illustrate this, we have computed the black-body-like spectrum with $b^\mu$ pure time-like, $b^\mu=(b^0\neq0; \vec{0})$. In this case, we get $\omega_{\pm}^{2}=c^2k(k\pm m)$, with $m=|b_{0}|$, so that:

\begin{equation}
u^{\rm odd}_{\rm timelike} (\omega_{\pm},T)|_{\omega>>\omega_0}= \frac{4\pi\hbar}{(2\pi c)^3}\frac{\omega^3}{e^{\hbar\omega/k_B T}-1} \left(1 \pm \frac12 \frac{\omega_0}{\omega} \mp \frac12 \frac{\omega^2_0}{\omega^2} +\frac12 \frac{\omega^2_0}{\omega^2} +{\cal O}(\omega^3_0)\right)\,. \label{Uoddtimelike}
\end{equation}
which implies in corrections similar the case where $\hat{k}\cdot\hat{b}=1$, $u^{\rm odd}_{\parallel} (\omega_{\pm},T)$.\\
\section{Conclusions and Prospects}

A study of the emission and absorption of photons by non-relativistic atomic electron, within a Lorentz-violating and CPT-odd or CPT-even electrodynamics, is presented. Our main results concern how Planck law is slightly modified whenever those violations are incorporated in the usual Maxwell electromagnetism. We have realized that for the CPT-odd case the deviations appear to be linear or quadratic in the rest-like frequency, $\omega_0=mc^2/\hbar$, according photon momentum and the background vector field are parallel or perpendicular each other. Actually, these ``corrections'' take place in the expressions for the energy density distribution for a given polarization (mode). If a thermal bath with photons equally populated by each mode concerns, then even in the case where $\hat{b}\parallel\vec{k}$ the correction appears to be quadratic in $\omega_0$, but a higher number of photons of $\omega_-$-mode should come than the another polarization. Such a difference in the occupation number show up linearly in $\omega_0$. At the limits of $T\to0$, $T\to\infty$, and $\omega=|\vec{k}|c\to\infty$ those corrections gives no contribution and all the energy densities per frequency, $U(\omega)$, recover their usual counterparts. In the situation where CPT-symmetry is kept the leading order deviation appears linearly in the violating parameter. This seems to be the best possibility, within our analysis, to search for possible Lorentz violation once CPT-even parameters are relatively high, compared with CPT-odd ones. For the CMB situation the power spectra would be diminished by a factor of the order of $\rho\sim 10^{-16}$.\\

As prospects for future investigation, we may quote the study of how such violations modify the analysis of the spin structure of photons. Indeed, in the CPT-odd situation, we have encountered several obstacles for carrying on this study, once even the {\em little group} of spatial rotations now depends upon the relative orientation of the photon momentum with the background field. In some cases, the little group seems to be reduced to the Abelian $SO(2)$-group \cite{HarritonPLA2007}, so that a quantization (discretization) of spin-like eigenvalues could be jeopardized.\\

Recently, Earth-based proposals to search for Lorentz violations have been emerged. Among other, some situations where radiation is confined may offer interesting possibilities, once violating effects may be enhanced by the parameters associated to the (small) size and geometry of the medium \cite{Mewes-cavity-PRD2007}. As it is well-known, some quantities associated to the electromagnetic radiation propagating in the interior of conducting wave-guides (cylindrical, for concreteness) increase as the cylinder radius is diminished. Once current dimensions lies at the nanoscale (for instance, fabricated by capping a carbon nanotube with a very thin conductor film \cite{nanotubewaveguide}), the small corrections associated to Lorentz violation can be, in principle, largely enhanced in these confined media \cite{workinpreparation}. Very recently, a similar apparatus have been proposed as a potential way to probe signals for non-linear Born-Infeld corrections to the Maxwell electrodynamics\cite{waveguideBI}.\\

\centerline{\large Acknowledgements\\}
\vskip .5cm
The authors are grateful to B. Altschull, R. Casana, M.M. Ferreira Jr., and J.A. Helay\"el-Neto for fruitful discussions and for having drawn their attention to some weak points in an earlier version of this manuscript. They also thank the Brazilian agencies CAPES, CNPq and FAPEMIG for financial support.

\end{document}